\documentclass[doublecol]{epl2}

\usepackage{amsmath}
\usepackage{amssymb}
\usepackage{graphicx}% Include figure files
\usepackage{bm}% bold math
\usepackage{epstopdf}

\newcommand{\prt}{\partial}
\newcommand{\sn}{\mathrm{sn}}

\newcommand{\om}{\omega}

\newcommand{\la}{\lambda}

\title{Periodic waves in two-component Bose-Einstein condensates with repulsive interactions between atoms}
\shorttitle{Waves in two-component condensates}

\author{A. M. Kamchatnov \thanks{E-mail: \email{kamch@isan.troitsk.ru}}}

\shortauthor{A. M. Kamchatnov }

\institute{
Institute of Spectroscopy, Russian Academy of Sciences, Troitsk,
Moscow, 142190, Russia
}

\abstract{
We consider periodic waves in miscible two-component Bose-Einstein condensates
with repulsive nonlinear interaction constants. Exact one-phase solution
is found for the case when all these constants are equal to each other
(i.e., for the Manakov limit). New types of nonlinear polarization waves are
considered in detail. The connection of the solutions found with
experimentally observed periodic structures in two-component condensates
is discussed.}

\begin{document}

\maketitle

\section{Introduction}
Possibility of relative motion between two species in two-component Bose-Einstein condensates (BECs)
leads to a rich phenomenology of nonlinear wave structures which can be generated in such systems.
For example, in a two-component BEC with different values of intra- and inter-atomic interaction
constants there exist two characteristic velocities of linear waves---the sound velocity of density
waves with in-phase motion of the two species and the polarization wave velocity with counter-phase
motion of the two species (see, e.g., \cite{ar-13}). Correspondingly, there are two Mach cones and two channels of Cherenkov
radiation of Bogoliubov linear waves by an obstacle moving with high enough speed \cite{susanto-07}. Then interference
of Cherenkov waves yields stationary ship wave patterns located outside the corresponding Mach cones.
Besides that, a supersonic flow past an obstacle can generate
oblique density solitons \cite{egk-06,gladush-09}, half-solitons \cite{fsm-11}, or oblique polarization breathers
\cite{kk-13}. These structures have been observed in experiments with flows of polariton condensates past
obstacles \cite{amo-2011,grosso-2011,half-sol-12}. In these experiments, both components had the same
incident flow velocity and the wave pattern was created due to the action of the obstacle's potential.
However, there exist other mechanisms of formation of nonlinear wave patterns. In particular,
such patterns can be created by development of instability in the system, and such a modulation
instability can be effective in the repulsive two-component condensates provided there exists large
enough relative velocity between the components \cite{law-01}. In the experiments
\cite{hamner-11,hoefer-11,hamner-13} it was demonstrated that the relative motion between two species leads at the nonlinear
stage of evolution to formation of the periodic polarization wave with locations of the crests in the density distribution of one component
coinciding with locations of the troughs in the density distribution of the other component, so that the total density remains practically
constant (although gradually changing along the trap at a larger scale of distance). These observations pose
the problem of finding periodic solutions of the Gross-Pitaevskii (GP)
equations describing dynamics of the two-component BECs. Although this problems was addressed in a number
of papers (see, e.g., \cite{pp-99,shin-04} and references therein), the general enough solution in the form
convenient for applications is still absent. In this Letter, we shall present such a solution and
discuss its possible applications to BEC dynamics.

\section{General solution}
In accordance with the experiments \cite{hamner-11,hoefer-11,hamner-13}, we consider a miscible two-component condensate
confined in a one-dimensional trap which in the first approximation can be considered as a uniform cylinder.
In the mean field approximation the dynamics of BEC can be described with a good accuracy by two-component GP equations
\begin{equation}\label{eq0}
\begin{split}
    &i\hbar\frac{\prt\Psi_+}{\prt t}+\frac{\hbar^2}{2m}\Delta\Psi_+-\left(g_{11}|\Psi_+|^2+g_{12}|\Psi_-|^2\right)\Psi_+=0,\\
    &i\hbar\frac{\prt\Psi_-}{\prt t}+\frac{\hbar^2}{2m}\Delta\Psi_--\left(g_{12}|\Psi_+|^2+g_{22}|\Psi_-|^2\right)\Psi_-=0,
    \end{split}
\end{equation}
where $\Psi_{\pm}$ are wave functions of the condensate's components and the nonlinearity constants $g_{ij}=4\pi\hbar^2a_{ij}/m$
can be expressed in terms of the scattering lengths $a_{ij}$. We suppose that the components consist of atoms in two different
electronic states and, hence, they have the same mass $m$. These components can mix if the constants of the nonlinear interaction
satisfy the condition $g_{12}^2<g_{11}g_{22}$ (see \cite{timm-98,ao-98}). In the experiments of refs.~\cite{hamner-11,hoefer-11,hamner-13}
these constants correspond to the states $|1,-1\rangle$ and $|2,-2\rangle$ of hyperfine structure of atoms $^{87}\mathrm{Rb}$ with
the scattering lengths equal to  $a_{11}=100.4a_0$, $a_{12}=98.98a_0$ and $a_{22}=98.98a_0$, where $a_0$ is the Bohr radius \cite{verhaar-09}.
Thus, they satisfy the above condition and, at the same time, they are very close to each other. Therefore for the
description of the dynamics of the condensate we can accept that these constants have the same value $g$. It is worth
noticing that in this case the velocity threshold for instability studied in ref.~\cite{law-01} vanishes, that is
an arbitrarily small relative velocity between the components leads to growth of the polarization wave.
In a rarefied enough condensate the transverse motion of atoms is described by the ground state function of a
two-dimensional oscillator. Averaging over this state reduces the system (\ref{eq0}) to the effectively one-dimensional
system which describes the axial motion of the condensate's components along the coordinate $x$. Then the effective
nonlinearity constant takes the value $g_{1D}=g/(2\pi a_{\bot}^2)$ where $a_{\bot}=(\hbar/m\om_{\bot})^{1/2}$ and $\om_{\bot}$ is
the frequency of the transverse oscillations of atoms in the trap. It is convenient to introduce non-dimensional variables
in the following way. Let $\rho_\ab{ch}$ be a characteristic density of atoms, say, at the trap center. Then we take the
healing length  $\hbar/\sqrt{mg_\ab{1D}\rho_\ab{ch}}$ as a unit of length, and the time $\hbar/(g_\ab{1D}\rho_\ab{ch})$ during which the
healing length is traversed by a wave propagating with the sound velocity $\sqrt{g_\ab{1D}\rho_\ab{ch}/m}$ as a unit of time.
If we introduce also the wave functions $\psi_{\pm}$ normalized according to $\Psi_{\pm}=\sqrt{\rho_\ab{ch}}\psi_{\pm}$,
then we arrive at the so-called Manakov system \cite{manakov}, where for convenience we use the previous notation
for the non-dimensional variables $x$ and $t$:
\begin{equation}\label{eq1}
i \partial_t \psi_{\pm} + \frac12 \partial_{xx}^2\psi_{\pm}
- \left(|\psi_\pm|^2+|\psi_{\mp}|^2\right) \psi_\pm =0.
\end{equation}
It is convenient to represent a two-component order parameter $(\psi_+(x, t),\psi_-(x, t))$
as a spinor variable \cite{ktu-2005}
\begin{equation}\label{eq2}
    \left(
            \begin{array}{c}
              \psi_+ \\
              \psi_- \\
            \end{array}
          \right)=\sqrt{\rho}\, e^{i\Phi/2}\chi=
          \sqrt{\rho}\, e^{i\Phi/2}
          \left(
            \begin{array}{c}
              \cos\frac{\theta}2\,e^{-i\phi/2} \\
              \sin\frac{\theta}2\,e^{i\phi/2}  \\
            \end{array}
          \right).
\end{equation}
Here $\rho(x,t)=|\psi_+|^2+|\psi_-|^2$ denotes the total density of the
condensate and $\Phi(x,t)$ has the meaning of the velocity potential of its
in-phase motion; the angle $\theta(x,t)$ is the variable describing the
relative density of the two components
($\cos\theta=(|\psi_+|^2-|\psi_-|^2)/\rho$) and $\phi(x,t)$ is the
potential of their relative (counter-phase) motion. Accordingly, the
densities of the components of the condensate are given by
\begin{equation}\label{eq2c}
    \rho_+(x,t)=\rho\cos^2(\theta/2),\quad \rho_-(x,t)=\rho\sin^2(\theta/2),
\end{equation}
and their phases are defined as
$\varphi_+(x,t)=\tfrac12(\Phi-\phi)$, $\varphi_-(x,t)=\tfrac12(\Phi+\phi)$.
The corresponding velocities of the components are equal to
\begin{equation}\label{eq2d}
v_{\pm}(x,t)=\prt_x\varphi_{\pm}=\frac12(U\mp v),
\end{equation}
where $U=\Phi_x$ and $v=\phi_x$.
Substitution of eq.~(\ref{eq2}) into eq.~(\ref{eq1}) yields the system \cite{kklp-13}
\begin{equation}\label{eq3}
    \begin{split}
    \rho_t&+\frac12[\rho(U-v\cos\theta)]_x=0,\\
    \Phi_t&-\frac{\cot\theta}{2\rho}(\rho\theta_x)_x
+\frac{\rho_x^2}{4\rho^2}-\frac{\rho_{xx}}{2\rho}\\
&+\frac14(\theta_x^2+U^2+v^2)+2\rho=0,\\
    \rho\theta_t&+\frac12[(\rho v\sin\theta)_x+\rho U\theta_x]=0,\\
    \phi_t&-\frac1{2\rho\sin\theta}(\rho\theta_x)_x+\frac12Uv=0.
    \end{split}
\end{equation}

We shall confine ourselves to situations when both components have equal
chemical potentials $\mu$ and the wave propagates with velocity $V$. Hence,
we look for the solution of the system (\ref{eq3}) in the form
\begin{equation}\label{eq4}
\begin{split}
    &\rho=\rho(\xi),\quad \theta=\theta(\xi),\quad \Phi=-2\mu t+\Phi_0(\xi),\\
    &\phi=\phi(\xi),\quad\text{where}\quad \xi=x-Vt.
    \end{split}
\end{equation}
In this case the system (\ref{eq3}) can be integrated and we shall describe
briefly the main steps of this calculation.
Substitution of eqs.~(\ref{eq4}) into the first equation (\ref{eq3}) and integration gives
\begin{equation}\label{eq5}
    \Phi_{\xi}-\cos\theta\phi_{\xi}=2(V-A/\rho),
\end{equation}
where $A$ is an integration constant. The function $\phi_{\xi}$ can be excluded from (\ref{eq5})
with help of the last equation (\ref{eq3}) where $\phi_t=-V\phi_{\xi}$. This gives the equation
for $\Phi_{\xi}\equiv\Phi_{0,{\xi}}$ whose solution reads
\begin{equation}\label{eq6}
    \Phi_{\xi}=2V-\frac{A+ K}{\rho},\quad\text{and, hence,}\quad
    \phi_{\xi}=\frac{A- K}{\rho\cos\theta},
\end{equation}
where
\begin{equation}\label{eq6a}
K=\pm\sqrt{A^2+\rho\cot\theta(\rho\theta_{\xi})_{\xi}}.
\end{equation}
The substitution of the expressions for $\Phi_{\xi}$ and $\phi_{\xi}$ into the third equation (\ref{eq3})
gives equation for $K$,
$$
K_{\theta}+(2\cot\theta+\tan\theta)K=-A\tan\theta,
$$
whose elementary integration yields
\begin{equation}\label{eq7}
    K=-A(2\cot^2\theta+1)+2B\frac{\cos\theta}{\sin^2\theta},
\end{equation}
where $B$ is an integration constant. Equating this to the expression for $K$ presented in (\ref{eq6a}), we get
the equation which can be integrated once to give
\begin{equation}\label{eq8}
    \left(\frac{d\theta}{d\xi}\right)^2=\frac4{\rho^2}\left(C^2-\frac{(B-A\cos\theta)^2}{\sin^2\theta}\right),
\end{equation}
where $C^2$ is an integration constant (it is clear from eq.~(\ref{eq8}) that it must be positive).
At last, the substitution of the obtained expressions for $\theta_{\xi},\,U=\Phi_{\xi},v=\phi_{\xi}$ into
the second equation (\ref{eq3}) where $\Phi_t=-2\mu-V\Phi_{\xi}$ leads to the equation which again
can be integrated once to give
\begin{equation}\label{eq9}
    \rho_{\xi}^2=4\mathcal{R(\rho)},
\end{equation}
where
\begin{equation}\label{eq10}
    \mathcal{R}(\rho)=\rho^3-(2\mu+V^2)\rho^2+D\rho-(A^2+C^2)
\end{equation}
and $D$ is one more integration constant.

It is remarkable that the variable $\rho$ is separated in eq.~(\ref{eq9}) from the other variables.
This is a consequence of the complete integrability \cite{manakov} of the Manakov limit (\ref{eq1})
of the two-component GP equations, although we have not used here this fact explicitly.
The solution of eq.~(\ref{eq9}) is parameterized by three
zeroes of the polynomial $\mathcal{R}(\rho)=(\rho-\rho_1)(\rho-\rho_2)(\rho-\rho_3)$, $\rho_1\leq\rho_2\leq\rho_3$,
and it can be expressed in standard notation in terms of the Jacobi elliptic function,
\begin{equation}\label{eq11}
    \rho(\xi)=\rho_1+(\rho_2-\rho_1)\sn^2(\sqrt{\rho_3-\rho_1}\,(\xi+\xi_0),m),
\end{equation}
where $m=(\rho_2-\rho_1)/(\rho_3-\rho_1)$. In this solution, the total density $\rho$ oscillates in the
interval $\rho_1\leq\rho\leq\rho_2$ and, according to its physical meaning, $\rho$ must be positive; hence
all $\rho_i$, $i=1,2,3,$ are positive, too. Their product denoted as $R^2\equiv\rho_1\rho_2\rho_3 $ equals to $R^2=A^2+C^2$.
Therefore, we can introduce, instead of the constants $A,\,B,\,C$, more convenient parameters $\beta$ and $\gamma$
as follows,
\begin{equation}\label{eq12}
    A=R\cos\gamma,\quad C=R\sin\gamma,\quad
    B=R\cos\beta.
\end{equation}
The last definition implies that $|B|\leq R=\sqrt{\rho_1\rho_2\rho_3}$ what follows from the observation that according
to eq.~(\ref{eq8}) the angle $\theta$ oscillates between the values $\theta_1,\,\theta_2$ determined by the condition
that the right-hand side of eq.~(\ref{eq8}) vanishes, that is by the equation $B=A\cos\theta_{1,2}\pm C\sin\theta_{1,2}=
R\cos(\gamma\mp\theta_{1,2})$ which gives $|B|\leq R$. We define
\begin{equation}\label{eq13}
    \theta_1=\beta+\gamma,\quad \theta_2=\beta-\gamma
\end{equation}
and suppose for definiteness that the parameters $\beta$ and $\gamma$ are chosen in such a way that
$\cos\theta_1\leq\cos\theta_2$. Then eq.~(\ref{eq8}) reduces to
\begin{equation}\label{eq14}
   \pm \frac{\sin\theta \upd\theta}{\sqrt{(\cos\theta-\cos\theta_1)(\cos\theta_2-\cos\theta)}}=2R\frac{\upd\xi}{\rho(\xi)}.
\end{equation}
Its integration yields the solution for $\theta(\xi)$:
\begin{equation}\label{eq15}
    \cos\theta(\xi)=\cos\theta_1\sin^2\frac{X(\xi)}2+\cos\theta_2\cos^2\frac{X(\xi)}2,
\end{equation}
where
\begin{equation}\label{eq16}
    X(\xi)=2R\int_{\xi_0}^{\xi}\frac{\upd\xi'}{\rho_1+(\rho_2-\rho_1)\sn^2(\sqrt{\rho_3-\rho_1}\,\xi',m)}+X_0,
\end{equation}
$X_0$ is an integration constant. The integral in (\ref{eq16}) can be expressed in terms of Weierstrass elliptic
functions (see, e.g., \cite{kamch-90}), however it is more convenient for future study to keep it in a
non-integrated form. Equations (\ref{eq11}) and (\ref{eq15}),(\ref{eq16}) determine the fields
$\rho(x,t)$ and $\theta(x,t)$. Their substitution into eqs.~(\ref{eq6}) and (\ref{eq2d}) yields
the flow velocities of the BEC components:
\begin{equation}\label{eq17}
    v_+=V-\frac{R}{2\rho}\cdot\frac{\cos\gamma+\cos\beta}{\cos^2(\theta/2)},\quad
    v_-=V-\frac{R}{2\rho}\cdot\frac{\cos\gamma-\cos\beta}{\sin^2(\theta/2)}.
\end{equation}
Subsequent integration of these formulae and account of the expression for the chemical potential
\begin{equation}\label{eq18}
    \mu=\tfrac12(\rho_1+\rho_2+\rho_3-V^2)
\end{equation}
gives the expressions for the phases $\varphi_{\pm}$ (see eqs.~(\ref{eq2d})). This completes the
derivation of the periodic solution of the system (\ref{eq1}). It is parameterized by
six constant parameters $V,\rho_1,\rho_2,\rho_3,\beta,\gamma$.

It is important to notice that only the total density $\rho(x,t)$ is a periodic function of $\xi=x-Vt$;
the angle $\theta$ is a periodic function of $X$ which is a quite complicated function of $\xi$
(see eq.~(\ref{eq16})). Therefore the densities $\rho_{\pm}$ of the components and their
velocities $v_{\pm}$ vary with $x$ and $t$ in a complicated way.
However, the situation greatly simplifies in important particular cases discussed below.

\section{Nonlinear polarization wave}
The spinor (\ref{eq2}) can be characterized by the polarization vector $\mathbf{S}=\chi^\dag \boldsymbol{\sigma}\chi$
(here $\boldsymbol{\sigma}=(\sigma_1,\sigma_2,\sigma_3)$ is a vector of Pauli matrices) with the components
$\mathbf{S}=(\sin\theta\cos\phi,\sin\theta\sin\phi,\cos\theta)$. Hence, the relative motion of two
components of BEC can be represented as a {\it polarization} dynamics. Here we shall consider the above
solution in the case of pure polarization dynamics when the total density is constant. This takes place
for $\rho_1=\rho_2\equiv\rho_0$, when $R=\rho_0\sqrt{\rho_3}$ and $\rho(\xi)\equiv\rho_0$. Hence we get
\begin{equation}\label{eq21}
    X(\xi)=2\sqrt{\rho_3}(\xi-\xi_0)+X_0.
\end{equation}
This is a linear function of $x$ and $t$, therefore such a polarization wave represented by eq.~(\ref{eq15})
depends periodically on the space and time variables. Instead of the parameter $\rho_3$, it is convenient to
introduce other parameters with clearer physical meaning. Let us define mean fluxes of the components
averaged over either $x$ or $t$: $j_{\pm}= \overline{\rho_{\pm}v_{\pm}}$. A simple calculation gives
\begin{equation}\label{eq22}
\begin{split}
    j_+&=\tfrac12\rho_0[V(1+\cos\beta\cos\gamma)-\sqrt{\rho_3}(\cos\beta+\cos\gamma)],\\
    j_-&=\tfrac12\rho_0[V(1-\cos\beta\cos\gamma)+\sqrt{\rho_3}(\cos\beta-\cos\gamma)].
    \end{split}
\end{equation}

Obviously, the amplitude of this wave can be measured by the parameter $(\theta_2-\theta_1)/2=\gamma$.
First, we shall consider linear waves propagating over a quiescent background with $\gamma=0$ and
$j_{\pm}=0$. The last condition is satisfied if $\sqrt{\rho_3}=V$. Then in the small amplitude limit
$\gamma\ll1$ eq.~(\ref{eq15}) takes the form
\begin{equation}\label{eq23}
    \cos\theta\cong\cos\beta+\gamma\sin\beta\cdot\cos[2V(x-Vt)].
\end{equation}
Here the wave number and the frequency are equal, respectively, to $k=2V$, $\omega=2V^2$, and they
satisfy the dispersion relation
\begin{equation}\label{eq23a}
    \omega(k)=\tfrac12k^2.
\end{equation}
This formula suggests that, on the contrary to Bogoliubov density waves, the linear polarization
waves can have arbitrarily small velocities what is confirmed by the direct analysis of the linearized
two-component GP equations---the ``sound'' velocity of the polarization waves vanishes in the Manakov
limit (see, e.g., \cite{kklp-13}). Hence, the parameter $\rho_3=V^2$ is not limited here by the inequality
$\rho_3>\rho_0$ implied above which is an artefact of our method of derivation based on the study of waves with changing
total density.

In a nonlinear wave both fluxes are equal to zero if either $\cos\beta=0$ or $\sqrt{\rho_3}=V$.
The first case corresponds to the condensates with equal mean densities of the two components. In this case we get
the nonlinear wave with
\begin{equation}\label{eq24}
    \cos\theta=\sin\gamma\cdot\cos\left[\frac{2V}{\cos\gamma}(x-Vt)\right],
\end{equation}
where the wave number and the dispersion relation depend on the amplitude $\gamma$:
\begin{equation}\label{eq25}
    k={2V}/{\cos\gamma},\quad \omega(k)=\tfrac12k^2\cos\gamma.
\end{equation}

The second opportunity $\sqrt{\rho_3}=V$ takes us back to the linear limit $\gamma=0$. This means that
in the condensates with nonequal mean densities the nonlinear waves can exist only if the fluxes
are not equal to each other, that is a counterflow is needed. This can be realized by a number of
different possibilities. Let us consider two typical examples.

Let the fluxes have equal magnitudes and opposite directions: $j_+=-j_-=-j_0$. Then from eqs.~(\ref{eq22})
we get
$$
 \sqrt{\rho_3}= \frac{V}{\cos\gamma}=\frac{j_0}{\cos\beta\sin^2\gamma}.
$$
This means that if $j_0>0$, then $\beta<\pi/2$, and vice versa. The wave number and the dispersion relation
are given by
\begin{equation}\label{eq26}
   k=\frac{2j_0}{\cos\beta\sin^2\gamma},\quad \omega(k)=\tfrac12k^2\cos\gamma.
\end{equation}

We can obtain a stationary wave with $V=0$ if $j_+^2\neq j_-^2$. In this case eqs.~(\ref{eq22}) yield
$$
\sqrt{\rho_3}= \frac{j_-^2-j_+^2}{4\cos\beta\cos\gamma}.
$$
Suppose that this wave is excited from a quiescent uniform state with ratio of the components densities
$\rho_-/\rho_+=\tan^2(\beta_0/2)$. If this parameter remains constant during the excitation process,
then the parameters in the excited wave satisfy the condition
\begin{equation}\label{eq27}
    \cos\beta\cos\gamma=\cos\beta_0.
\end{equation}
Hence, the wave number of the stationary wave equals to
\begin{equation}\label{eq28}
    k=2\sqrt{\rho_3}=\sqrt{\left|\frac{j_-^2-j_+^2}{\cos\beta_0}\right|}.
\end{equation}

\begin{figure}[ht]
\begin{center}
\includegraphics[width=6.5cm]{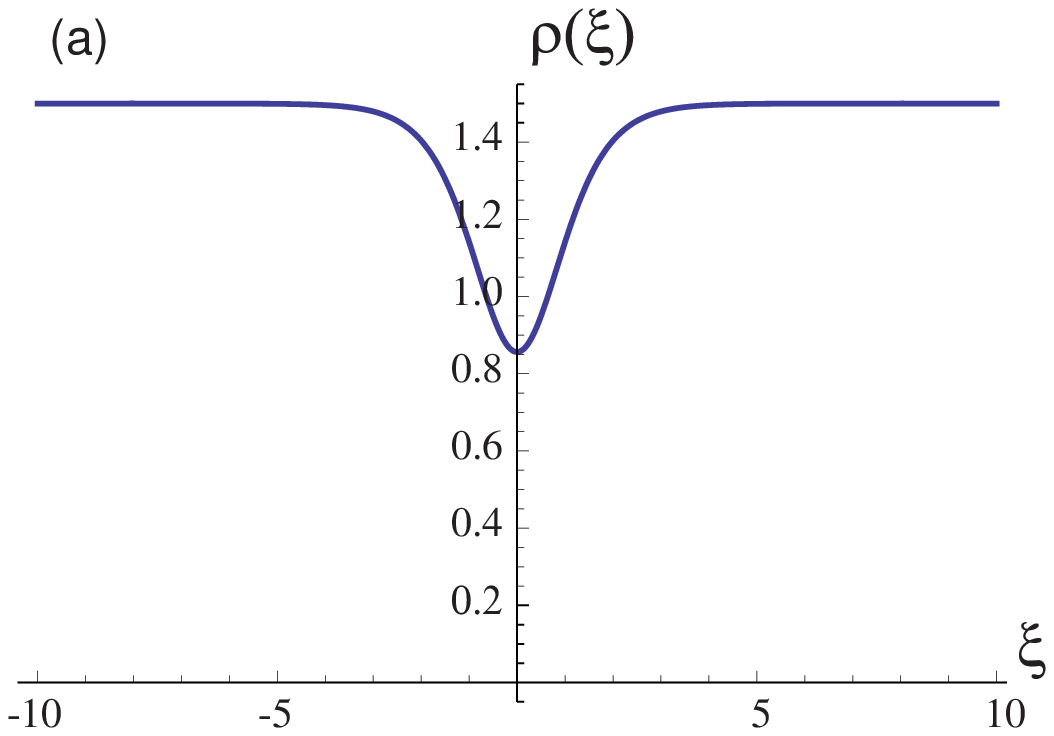}\\
\includegraphics[width=6.5cm]{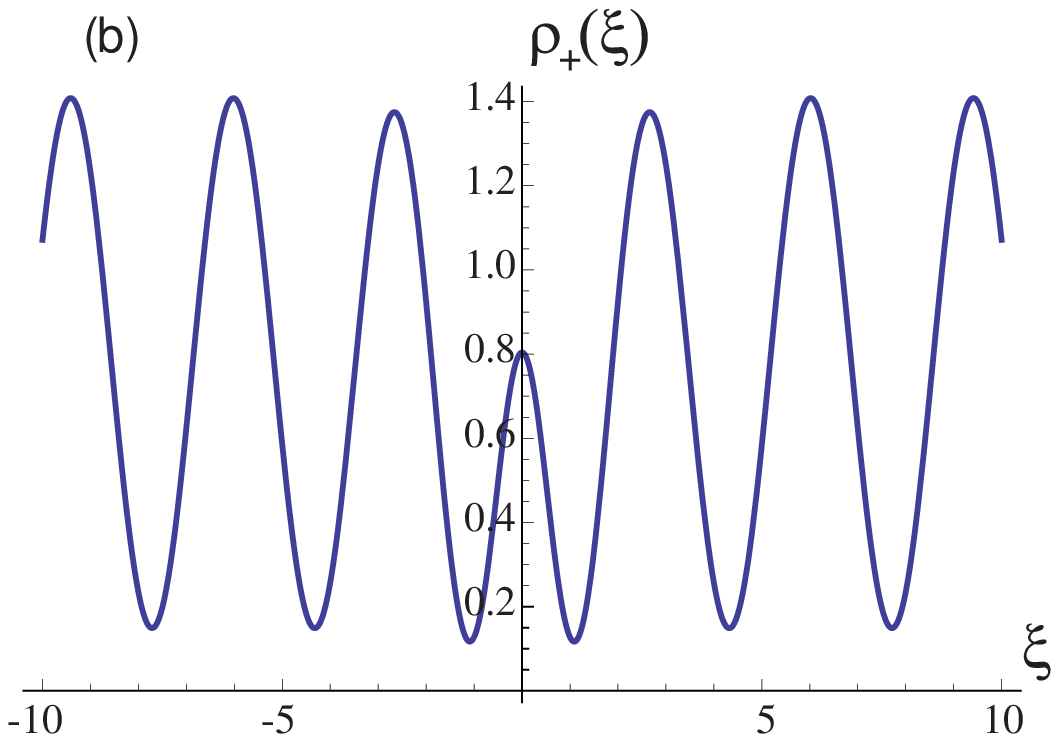}\\
\includegraphics[width=6.5cm]{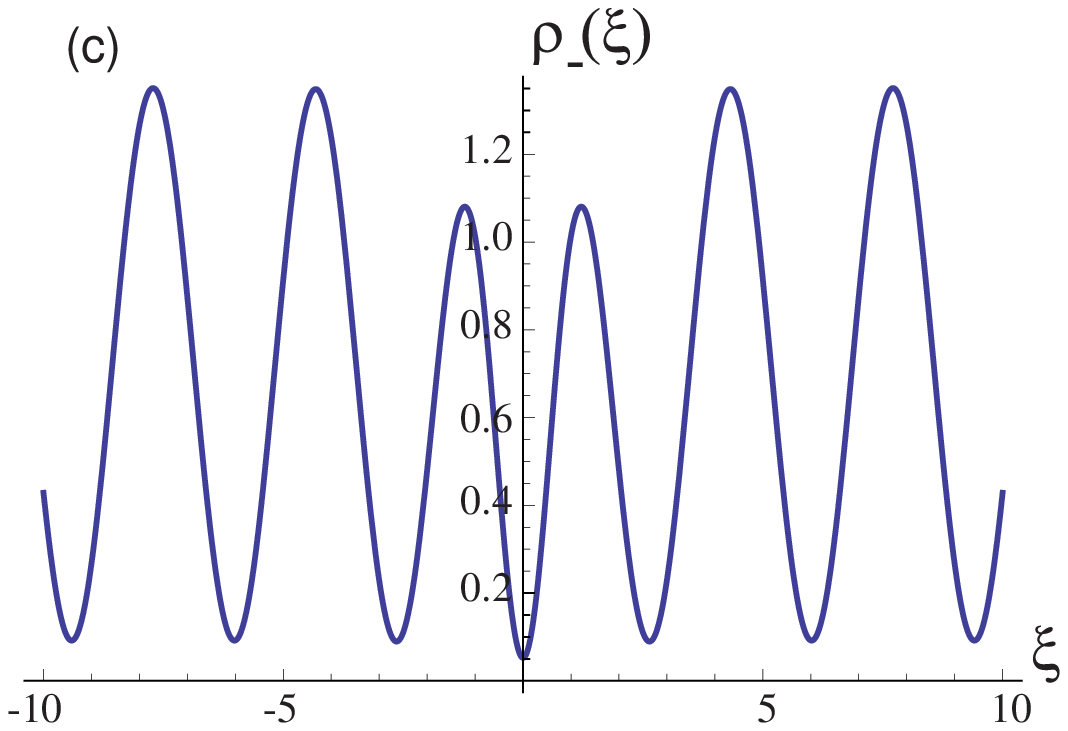}
\caption{Dependence of the total density $\rho(\xi)$ (a) and the components densities $\rho_+(\xi)$ (b), $\rho_-(\xi)$ (c) on $\xi$.
The parameters of the wave are equal to $V=0.5$, $\rho_0=1.0$, $\rho_1=0.857$, $\beta=1.5$, $\gamma=1.0$.
}
\end{center}
\label{fig1}
\end{figure}

\section{Quasi-soliton wave}
Now we shall turn to the soliton limit of the solution (\ref{eq11}) for the total density when $\rho_2=\rho_3=\rho_0$ and
this solution takes the form
\begin{equation}\label{eq29}
    \rho(x,t)=\rho_0\left\{1-\frac{1-\rho_1/\rho_0}{\cosh^2[\sqrt{\rho_0-\rho_1}(x-Vt)]}\right\}
\end{equation}
(to simplify the notation, we suppose that the soliton is located in our reference frame at the point $x=Vt$). Far enough from the
soliton's center, the total density becomes uniform and equal to $\rho\cong\rho_0$. However, for $\theta_1\neq\theta_2$,
the components densities oscillate here and compose the polarization wave discussed above. The variable $X(x,t)$
defined by eq.~(\ref{eq16}) can be expressed now in terms of elementary functions,
\begin{equation}\label{eq30}
    X(\xi)=2\sqrt{\rho_1}\,\xi+2\arctan\left[\sqrt{\rho_0/\rho_1-1}\tanh(\sqrt{\rho_0-\rho_1}\,\xi)\right]
\end{equation}
(for simplicity, we omitted here $X_0$). It is clear that at $|\xi|\to\infty$ the variable $X$ becomes a linear
function of $\xi$. If we demand that at infinity both fluxes $j_{\pm}$ are equal to zero, then we obtain
\begin{equation}\label{eq31}
    \sqrt{\rho_1}=V/\cos\gamma
\end{equation}
Thus, this {\it quasi-soliton} wave transforms here into ``slow'' polarization waves with the parameter
$\rho_1<\rho_0$ playing the role of $\rho_3$ in the formulae of the preceding section.
(This confirms our statement above that the parameters in the polarization wave are not
limited by the condition $\rho_3>\rho_0$.) We call this solution a  {\it quasi-soliton} because it is
not localized in space on the contrary to the usual soliton solutions of nonlinear wave equations;
rather, generally speaking, it represents a ``defect'' in the polarization wave.
This defect manifests itself as a dip in the distribution of the total density in the form of a dark soliton.
The whole structure propagates with velocity $V$.
Equation (\ref{eq31}) can be rewritten in the form
\begin{equation}\label{eq32}
    V=V_s\cos\gamma,
\end{equation}
where $V_s=\sqrt{\rho_1}$ is the soliton's velocity related with the minimal density $\rho_1$ at its
center in the case of a one-component condensate. Thus, a quasi-soliton propagates slower than a
usual dark soliton with the same depth. When the amplitude of the polarization
oscillations vanishes, $\gamma=0$, then the quasi-soliton transforms into a well-known localized
dark-dark soliton with constant ratio of the components densities (see, e.g., \cite{gladush-09}).
\begin{figure}[ht]
\begin{center}
\includegraphics[width=6.5cm]{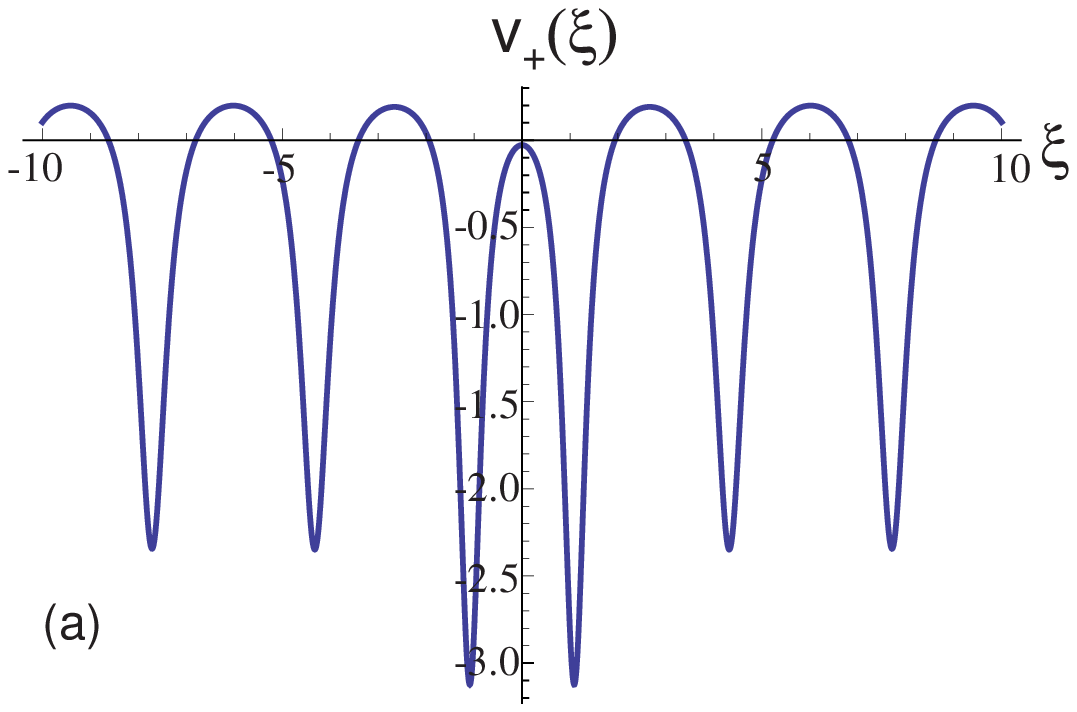}\\
\includegraphics[width=6.5cm]{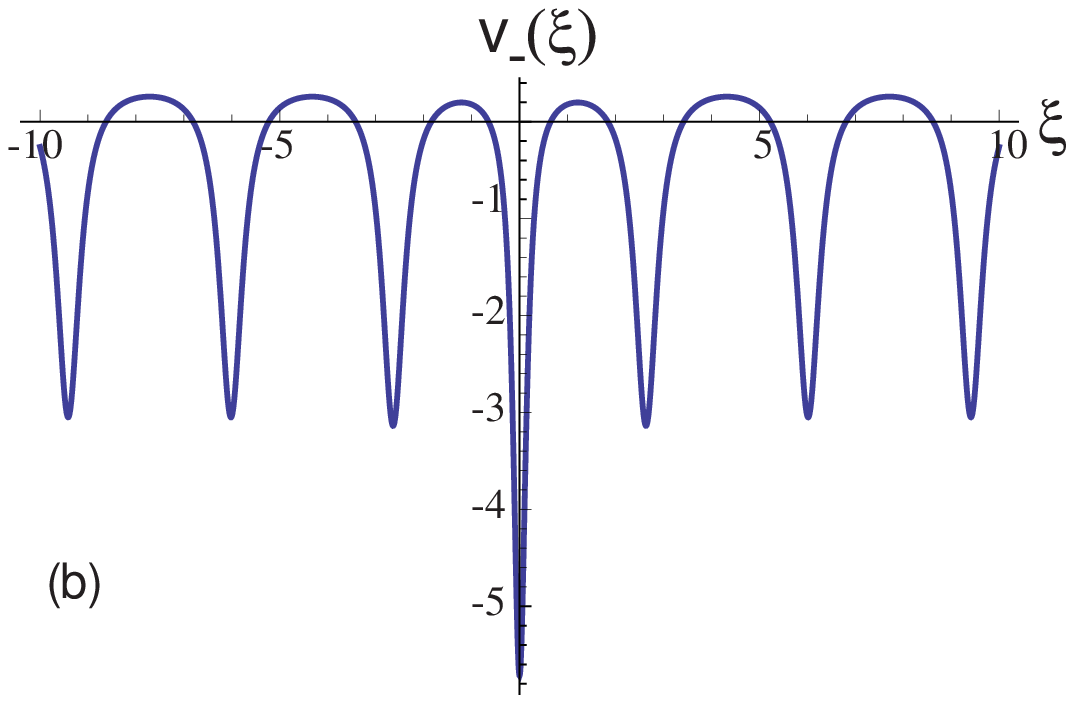}
\caption{Dependence of the flow velocities $v_+(\xi)$ (a) and  $v_-(\xi)$ (b) on $\xi$
of the condensate components.
The parameters of the wave are the same as in fig.~1.
}
\end{center}
\label{fig2}
\end{figure}

The distributions of the total density and the components densities in the quasi-soliton solution
are illustrated in fig.~1. It is clearly seen that far enough from the soliton center the
densities oscillate in counter-phase resulting in the constant total density. The distributions of the
flow velocities in this wave are shown in fig.~2. They show that the relative motion
between the components is crucially important for formation of such a wave.

\section{Discussion}

One can suppose that this type of a polarization wave has been observed in the experiments \cite{hamner-11,hoefer-11,hamner-13}.
In refs.~\cite{hamner-11,hoefer-11} the relative motion between the components of the two-state $^{87}$Rb condensate was induced by
a gradient of an external magnetic field applied along the axial direction of a cigar-shaped optical dipole trap. In \cite{hamner-11},
apparently a dispersive shock wave represented by a modulated polarization wave has been observed with large-amplitude oscillations
of the polarization and very small oscillations of the total density. In \cite{hoefer-11} a dense lattice of the polarization oscillations
has been created by a counterflow through the condensate due to the modulational instability of a uniform state \cite{law-01}. We believe
that the polarization waves, whose theory has been developed here, can be formed at a nonlinear stage of evolution of the modulational instability.
To support this supposition, we shall make here a rough theoretical estimate of the parameters of the wave and compare it with the
experimental data.

The main qualitative result of the theory is the statement that the wave number of the polarization wave is proportional to the velocity
of the wave, $k\propto V$, where the proportionality coefficient depends on the amplitude of the wave (see, e.g.,
(\ref{eq24}) or (\ref{eq26})). The dependence of $k$ on the parameters $\beta$ and $\gamma$ is weak; one can find that
it introduces an essential correction (a factor about 2) if only the density of one of the components is less than 20\% of the
density of the other component. To transform the approximate relation
$$
k=\frac{2\pi}{\la}\approx 2{V}
$$
($\la$ is the wavelength) to a dimensional form, we
calculate the healing length $\xi_{\mathrm{1D}}=\hbar/\sqrt{mg_\ab{1D}\rho_\ab{ch}}\cong2.4\times 10^{-5}\un{cm}$ and the sound velocity
$c_\mathrm{s}=\sqrt{g_\ab{1D}\rho_\ab{ch}/m}\cong0.3\un{ cm\cdot {s}^{-1}}$.
As a result we get the formula
\begin{equation}\label{eq48}
    \la\approx\frac{\pi\xi_\ab{1D}c_\mathrm{s}}V.
\end{equation}
From the data presented in \cite{hamner-11} one can find that the velocity of the polarization wave equals to
$V\cong1.8\times10^{-2}\un{cm\cdot s^{-1}}$. Substitution of these experimental parameters into eq.~(\ref{eq48})
gives $\la\approx13\un{\mu m}$. According to the experimental plots presented in \cite{hamner-11},
the experimental value of the wavelength equals to $\la\cong15-18\un{\mu m}$. Thus, taking into account
rough approximations which we have made in our estimate, we can consider this as a satisfactory agreement
of the theory with the experiment.
For more accurate theoretical description of formation of the dispersive shock wave generated in the experiment,
one needs to develop the theory of time-dependent modulations (Whitham theory) which can be the subject of
a separate study.

\section{Conclusion}
We have found an exact analytical periodic solution of the GP equations in the Manakov limit corresponding to equal constants
of repulsive interactions between atoms. This model can be considered as an approximation to description
of dynamics of a ``weakly miscible'' two-component condensate \cite{hamner-11,hoefer-11,yan-12}. In
immiscible two-component condensates other types of solitons generated by counterflow between the two
components are possible \cite{malomed-13}.
The solution found here includes as limiting cases the nonlinear polarization waves and quasi-solitons which
can be thought of as defects in the polarization waves.
In the limit of vanishing polarization oscillations, a quasi-soliton transforms to a standard dark
Manakov soliton. Existence of new solutions of the Manakov system, which can be considered as generalizations
of one-component dark solitons, poses the important problem of their stability. One may hope that the methods of
previous studies (see, e.g., \cite{barashenkov, pka-96}) can be generalized to this new situation, although
the problem of stability of polarization waves and quasi-solitons is far beyond the scope of this Letter.
We indicate only that possible relation of the polarization waves with experimentally observed wave
patterns suggests their stability.

In conclusion we remark that the theory developed here can be also applied to description of the polarization dynamics of light pulses in
nonlinear optical systems and the method used here can be extended to other choices of signs of the nonlinear interaction
constants.

\end{document}